# Monte Carlo simulation of magnetic phase transitions in Mn doped ZnO

L.B Drissi[1,2] *, A. Benyoussef [1,3], E.H Saidi[1, 4, 5], M. Bousmina[1]

1. INANOTECH, Institute of Nanomaterials and Nanotechnology (MAScIR), Rabat, Morocco,
2. International Centre for Theoretical Physics, ICTP, Trieste, Italy
3. Lab Magnetisme et PHE, Faculté des Sciences, Univ Mohammed V, Rabat, Morocco,
4. LPHE, Modelisation et Simulation, Faculté des Sciences, Univ Mohammed V, Rabat, Morocco
5. CPM, Centre of Physics and Mathematics-Rabat, Morocco.

**Abstract**

The magnetic properties of Mn-doped ZnO semi-conductor have been investigated using the Monte Carlo method within the Ising model. The temperature dependences of the spontaneous magnetization, specific heat and magnetic susceptibility have been constructed for different concentrations of magnetic dopant Mn and different carrier concentrations. The exact values of Mn concentration and carrier concentration at which high temperature transition occurs are determined. An alternative for the explanation of some controversies concerning the existence and the nature of magnetism in Mn diluted in ZnO systems is given. Other features are also studied.

---

* e-mail : ldrissi@ictp.it

# Introduction

In the recent years, a great interest has been devoted to Diluted Magnetic Semiconductors (DMS) doped with a small concentration of magnetic impurities inducing ferromagnetic DMSs. In particular, DMS based on III-V and II-VI semiconductors doped with transition metal are deeply investigated by both theoretical and experimental scientists [1-11] in order to use them for spintronic devices such as spin-valve transistor, spin light emitting diodes, optical isolator, non-volatile memory.

Theoretical studies have revealed interesting results. Dietl et al. [1] predicted high temperature ferromagnetism in some p-type doped semiconductors such as ZnO, GaN, GaAs and ZnTe. Later, this room temperature ferromagnetism was reported for the case of polycrystalline $Zn_{1-x}Mn_xO$ samples in [2] and [3]. Consequently, this oxide DMS has become one of the promising candidates for room temperature ferromagnetic DMSs.

ZnO based DMSs offer other desirable features as a semiconductor host. It has a direct wide-band gap (~3.4 eV), large exciton binding energy at room temperature (~ 60 meV). Ferromagnetic doped ZnO is a transparent piezoelectric ferromagnet with promoting applications in spintronics.

The study of these magnetic systems features by numerical simulation techniques has attracted great attention. This has, lead to several results. The ferromagnetic properties have been predicted in the work [4] using First-principle calculations and proved later by several experiments [5-9].

The incorporation of Mn into ZnO not only causes the introduction of magnetic moments but also increases the band gap. With co-doped concentration, the band gap decreases because a large hybridization between *p-d* orbitals was observed [6], so the *p-d* exchange not only plays the key role in the ion-ion ferromagnetic d-d exchange but also decreases the band gap.

From experimental point of view, there is a controversy concerning the existence and the nature of magnetism in Mn-doped ZnO systems. There exists some data where no ferromagnetism has been found [12]. While there is also other data on many systems which exhibit ferromagnetism with Curie temperature higher than room temperature [13]. However, this ferromagnetism is non conventional and seems to be a strange sort of 'intrinsic' ferromagnetism [14, 15]. Coey [16] has shown that the origin of the ferromagnetism is intrinsic; it is due to defects [17]. Indeed, sample without defects are paramagnetic, while those containing defects may be ferromagnetic. Ab initio calculations of transition metal doped ZnO, without defects, show that the ferromagnetic phase is stable (except for Mn, which is paramagnetic) [18]. Using LDA functional, the paramagnetic phase in Mn-doped ZnO has been confirmed by introducing self interaction correction, however, it has been stabilized for other transition metal doped ZnO [19]. The necessity to insert carriers, especially holes, into the system to stabilize the ferromagnetic phase was reported in details in [1]. Moreover, codoping (Zn, Mn)O system with N can change the ground state from spin glass to ferromagnetic as investigated in [17]. Depending on the functional used, the DFT results seem also controversial as showed in details in the two recent works [20, 21] for the case of Co-doped ZnO that is very well investigated in the literature[1].

Additionally, these conflicting results also exist concerning the distribution of Mn in ZnO as reported in the experimental results of [22] where Mn is distributed homogeneously and in [23] that reports clustering of Mn atoms or in [24] that studies dense nanograined Mn doped ZnO polycrystals. Thus, inconsistent phenomena and conclusions have been obtained, showing that the intrinsic ferromagnetism of Mn-doped ZnO systems remains an open question. It is hence of vital importance to clarify the correlation, if any, between carriers and the mechanism of ferromagnetism inherent to this class of diluted magnetic oxides.

---
[1] The authors thank the reviewer for pointing out this idea.

To overcome these discrepancies, ZnO have been well investigated employing ab-initio calculations as reported previously [4], [18-25], however Monte Carlo simulations are rarely examined for general diluted semiconductors, such the work highlighting the study of Co [26], and we are not aware of any studies for Mn-doped ZnO. In this work, we study in the framework of Monte Carlo simulations ZnO diluted by Mn to shed more light on the ambiguities concerning reports of ferromagnetism. In our model, the hole concentration is considered globally and it is introduced in the RKKY coupling via Fermi wave number as it is given explicitly in section 2. Amongst our results, spontaneous magnetization, specific heat, magnetic susceptibility and the Curie temperature $T_C$ have been evaluated for different concentrations of magnetic impurities and carriers of Mn-doped ZnO. This leads us to study the effect of both magnetic impurities and hole carriers on the existence and nature of ferromagnetism in $Zn_{1-x}Mn_xO$ and to give the values of concentration of carriers that could perform its magnetic order and those that must be avoided. Other results are also investigated.

## Model description

ZnO has wurtzite crystal structure (P63mc), where each atom of Zinc is surrounded by four cations of oxygen at the corners of a tetrahedron and vice versa. This tetrahedral coordination is typical of $sp^3$ covalent bonding and some of the divalent sites Zn substituted by the magnetic ions Mn. In the wurtzite structure, the two lattice constants of the hexagonal unit cell are $a=3.27A$ and $c=5.26A$ as reported for (Zn,Mn)O in [27].

ZnO system is n-type doped with free electrons in the conduction band or p-type with free holes carriers in the valance band. Based on the Hund's rule, we know that $Mn^{2+}$ with his half filled 3d shell has $S = 5/2$ ground state. So $Mn^{2+}$ introduce $d$ levels in the band gap of ZnO semiconductor According to the Zener model approach, ferromagnetism in ZnO originates from the RKKY–like interaction between the localized Mn spins via the delocalized holes carriers' spins [1]. The system is described by the Hamiltonian

$$\mathbf{H} = \sum_{i,j} J(r_{ij})\left(n_i n_j S_i^Z S_j^Z\right) \qquad \text{for } i \neq j \qquad (1)$$

where $r_{ij}$ is the separation between moments at the two sites $i$ and $j$ in the hexagonal structure and $n_i$ is the magnetic impurity occupation number. In H, the RKKY range function $J(r_{ij})= J^{RKKY}(r)$ is given as follows

$$J^{RKKY}(r) = J_0 e^{\frac{-r}{l}} r^{-4} \left[\sin(2K_F r) - 2K_F r \cos(2K_F r)\right] \quad (2)$$

where the Fermi wave number $K_F = (\frac{1}{3}\pi^2 n_c)^{1/3}$ depends on the hole density $n_c$. The positive coefficient $J_0$ is related to the local Zener coupling $J_{pd}$ between the Mn local moments and the hole spins. As we are concerned with short ranged RKKY coupling, we assume that the damping scale $l$, in the damped factor $e^{\frac{-r}{l}}$, is exactly the distance $D_1$ between the first nearest neighbors. For long range, the scale $l$ corresponds to the distance $D_3$ between the third nearest neighbors. We employ the above approach to estimate magnetic interactions in DMSs [28], [29].

For a site i, the spin $S_i^z$ introduced in eq(1) takes the values $\pm 1/2, \pm 3/2, \pm 5/2$. The corresponding energy $E_i$ is defined as follows:

$$E_i = n_i S_i^z \sum_j J^{RKKY}(r_{ij}) n_j S_j^z \quad . \qquad (3)$$

In our calculations, the concentration $n_i$ of Mn dopants in ZnO takes values in the interval $[0.15\text{-}0.35]$, as solubility of Mn in the ZnO matrix is relatively high ($x \leq 0.35$) as showed in [30] by pulsed laser deposition (PLD). Moreover, it was reported in [31] that co-doping (Zn, Mn)O system with N can change the ground state from spin glass to ferromagnetic. For p-type ZnO, the use of co-doping method could pave the way for a promising potential of Mn doped ZnO but this depends on the holes mediated ferromagnetism. This is why we are interested, in the current work, in investigating the carrier-induced ferromagnetism in the ZnO-based DMSs. So notice that the hole concentration that we consider globally are introduced in the RKKY coupling via Fermi wave number (eq. 2). following some previous works such as [31] that gives a generalized RKKY description of DMS, and going beyond [33] that uses effective field theory, with a Honmura-Kaneyoshi differential operator technique, to calculate the transition temperature as a function of the carrier (hole) $n_c$ and impurity $n_i$ concentrations for p-type ZnO diluted magnetic semiconductors. Recall that the Hamiltonian employed there contains a damped and undamped Ruderman-Kittel-Kasuya-Yosida (RKKY) interaction model to describe the exchange coupling constants $J_{ij}$ between the local moments $Mn_i$ and $Mn_j$. The method used is an effective field theory taking into account self correlations, but still neglecting correlations. It is well known that this approximation over estimates the transition temperature. As we mentioned before, in this present work we will go beyond this approximation, by using Monte Carlo simulation which gives more precise results. In this sense, our results are more precise than those obtained by effective field theory [33]. Indeed, as it is showed in the following sections, for [0- 0.2] we obtained lower transition temperature. Moreover, we investigated a larger domain of parameter than that investigated in [33]; in the region [0.52- 0.65] we obtained highest value of $T_C$, while in the intermediate region [0.2- 0.52] the system is nonmagnetic.

## Computational details and Results

We perform Monte Carlo simulations for the Ising model described above. The supercell, employed in these calculations, consists of 319-atoms, namely it represents $2\times2\times2$ primitive unit cell. The periodic boundary condition is applied. We have also used the supercells $3\times3\times3$ and $4\times4\times4$, which corresponds respectively to 612 and 1270 atoms, to determine the size effect. A specific number of nonmagnetic Zn atoms are replaced by the Mn impurities randomly.

The calculations need some input. As a first step, we determine the nearest neighbors. We restrict our calculations to the third first ones to take account of competition between ferromagnetic and antiferromagnetic interactions. Then, we use eq(2) to compute $J(r_{ij})$ for each distance $r_{ij}$. This step is crucial as this RKKY range function is required to calculate the total energy E of the system by using eq (3). It is also important to determine both the values of the hole density $n_c$ for which strong ferromagnetic coupling take place and the values corresponding to frustration phenomena as depicted in figure1.

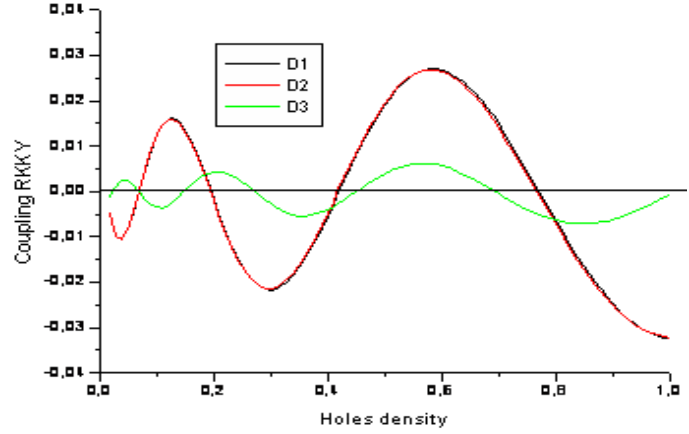

**Figure 1**: *Variation of RKKY function J($r_{ij}$) in terms of hole density $n_c$ for the three distances D1, D2 and D3 of the third first nearest neighbors*

With these ingredients at hand, we build our program where the thermal average of magnetization M and energy E are calculated by means of the Metropolis algorithm [34]. In this algorithm the Monte Carlo steps are 20000 per site using every 10-th step for averaging. We average over at least 100 different random configurations of magnetic sites of the disordered system for the quenched disorder average.

In this current work, we study the series of the system $Zn_{1-x}Mn_xO$. The data got from Monte Carlo simulations (MC) lead to the following results. We get the magnetization M and the energy E as function of temperature for different magnetic cations concentrations, $n_i$, and hole concentrations, $n_c$, for short ranged RKKY interaction. In figure2 we display the magnetization M versus temperature for a fixed carrier concentration $n_c$=0.18 and varied Mn impurity concentrations $n_i$. When T increases the magnetization M decreases and goes to zero. And it increases as $n_i$ increases 0.15, 0.25, 0.3 and 0.35 which is due to the alignment of spins that are antiferromagnetic as it is clearly shown in figure1 for $n_c$=0.18.

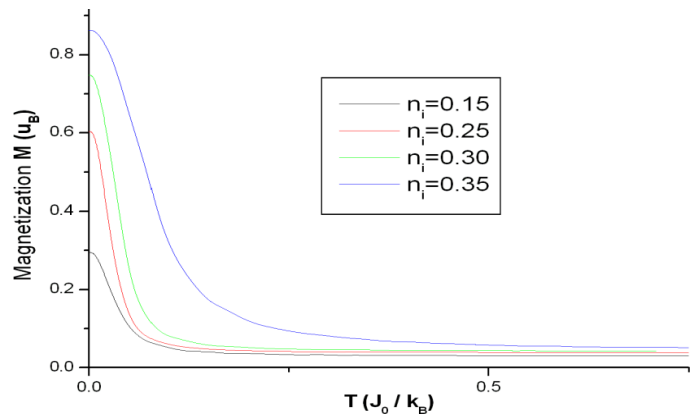

**Figure 2**: *Magnetization versus temperature for some different impurity concentrations $n_i$ , at $n_c$=0.18*

We also get the temperature dependence of the susceptibility. Indeed, correspondingly to M, the associated magnetic susceptibility $\chi \approx \frac{1}{T}(\langle M^2 \rangle - \langle M \rangle^2)$ is represented in figure 3 for $n_i$=0.2, 0.25 and 0.3. The three curves peak for different values of T that decreases as the concentration $n_i$ reduces. These computations are done as previously for a fixed carrier concentration $n_c$=0.18. The $\chi(T)$ peaks suggest that thermal fluctuation are driven phase transition from FM to paramagnetic that is in agreement with standard results.

From the three graphs of figure 3, we estimate the Curie temperatures of Mn-doped ZnO as a function of the Mn substitutional concentration $n_i$ for the specific $n_c = 0.18$. Notice that we plot here the results only for $n_c = 0.18$, however we have done the calculations for different values of $n_c$ as we will show in what follows.

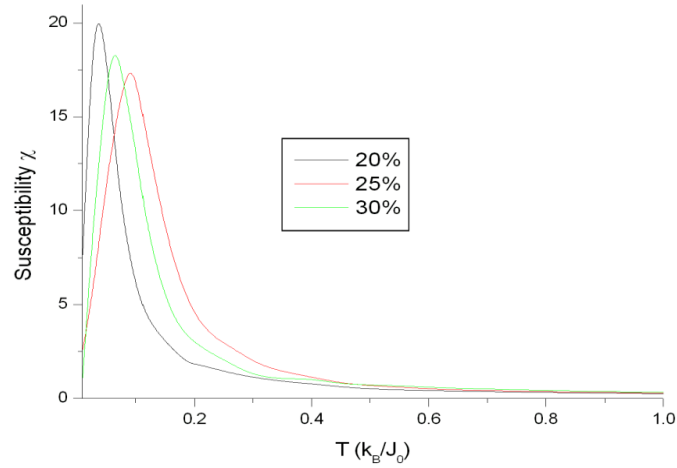

**Figure 3**: *Temperatute dependence of susceptibility for the 3 magnetic impurity concentrations $n_i$ 0.20, 0.25 and 0.3 at $n_c$=0.18*

To focus in this direction, we vary values of $n_i$ in the large interval [0.2-0.9] for the above mentioned and other holes density values $n_c$, namely 0.14, 0.18 and 0.55, 0.64, the obtained results of this study are presented here below in figure 4 where $T_C$ is depicted vs different values of $n_i$ for some fixed $n_c$. One sees $T_C$ increasing linearly with impurity concentration for the choosen four values of $n_c$. Generally, this behavior of $T_C$ is similar whatever $n_c$ is. However, from figure 4, it is obvious that hole density has also an effect. To understand it let us project the four values of $n_c$ in figure 1 that describes the short ranged RKKY coupling. We learn that, as $n_i$ is fixed, the highest and lowest transition temperature values depends on the nature of coupling. So, it is high in the region where the third first nearest neighbors coupling are ferromagnetic and it is low where only the first and second nearest neighbors coupling are ferromagnetic (see figure 1 for more details).

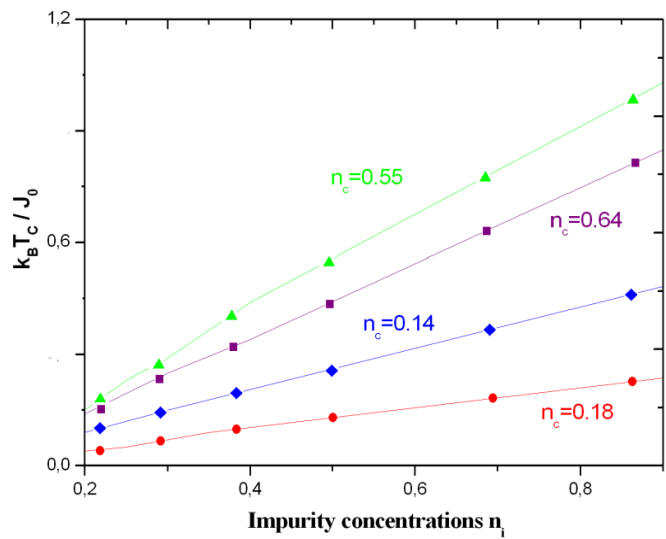

**Figure 4**: *The transition temperature, $T_c$, as function of the impurity concentrations $n_i$ for four different values of $n_c$.*

More information explaining the choice of these numbers will be given here below. At fixed $n_c = 0.18$, $T_C$ as function of Mn concentrations but for different range; namely short and long range that correspond respectively to damping scale $l = D_1$ and $D_3$, shows no deviation from strictly linear behavior as plotted in figure 5.

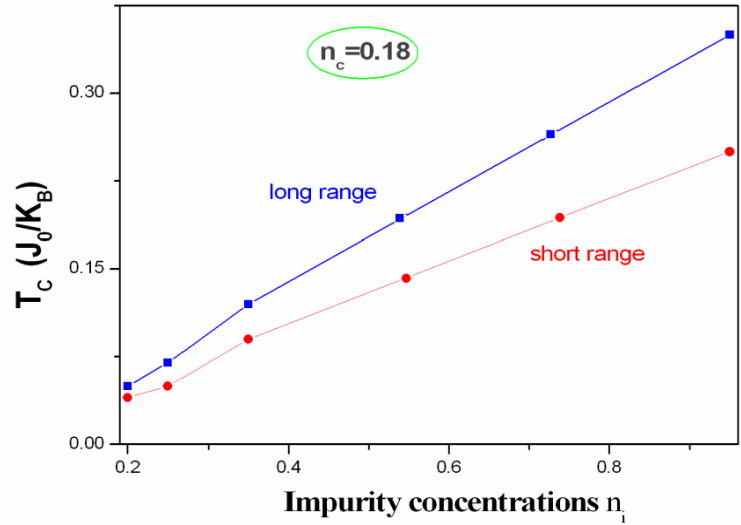

**Figure 5:** *the Curie temperature versus the impurity concentrations $n_i$ for the short and long range case, for $n_c=0.18$*

Since the transition temperature increases monotonically with magnetic impurities concentration $n_i$ for both long and short ranged RKKY coupling, then we will concentrate on the effect of carriers as well as the size of the cell on the variation of $T_C$ for fixed values of Mn concentrations. This is accomplished through collecting data from temperature dependence susceptibility by fixing the impurity concentration at a fixed value $n_i$ and varying the holes density. For the case[2] $n_i = 0.30$, we get the variation in the $T_C$ of $Zn_{0.7}Mn_{0.3}O$ versus carriers concentration $n_c$ as shown in figure 6.

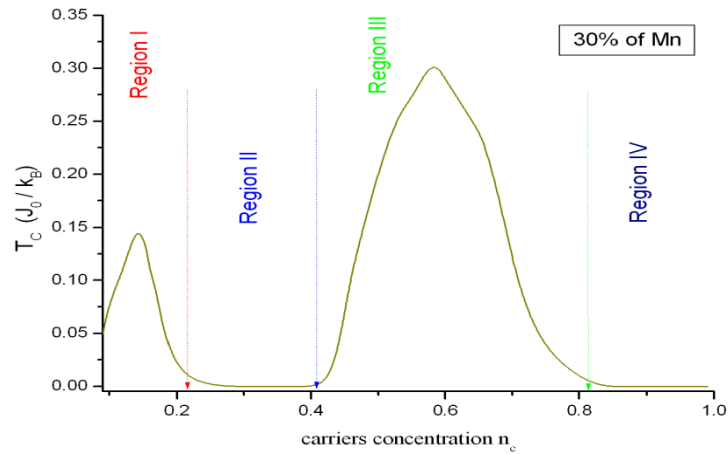

**Figure 6:** *The transition temperature, $T_c$, as function of the carriers concentration, $n_c$, for the specific Mn concentration 30%.*

The evolution of the Curie temperature in this figure reveals that we have four main different regions. Region I where $T_C$ peaks for $n_c=0.14$ and then decline. After, the transition temperature vanishes for $n_c$ taking values in the interval $[0.21\text{-}0.42]$ that coincides with Region II. In this last region we have competition between ferromagnetic and antiferromagnetic coupling that destabilizes ferromagnetism, leading to frustration. After this region the Curie temperature increases about 2 times as compared to that for low carrier concentrations (namely Region I) to maximize for $n_c=0.58$ and declines to vanish, for $n_c \geq 0.82$ that corresponds again to the frustrated Region IV. In order to shed more light on this data, it is necessary to focus on (eq 2). According to the study of the variation of $J^{RKKY}(r)$ in term of $n_c$ as showed in figure 1, we learn that region I corresponds to the case where the first and second near neighbors are ferromagnetic and the third one is antiferromagnetic while in region III all the three neighbors are FM. On the other hand, in both region II and IV the frustration is due to the dominant of AF mechanism. This confirms that in order to stabilize the FM phase in $Zn_{1-x}Mn_xO$ system, it is necessary to insert carriers with concentration $n_c$ belonging to the regions I or much better to region III. So, the various experimental and theoretical investigations of the magnetic order in $Zn_{1-x}Mn_xO$ doesn't give contradictory results. The ferromagnetism in (Zn,Mn)O systems reported in [2] or the anti-ferromagnetic or spin-glass behavior observed in [12], [17] could be now explained. The region I and region III agree with the results exhibiting ferromagnetism with Curie temperature higher than room temperature [13]. While the region II and IV correspond both to the experimental study [35] showing there is no evidence for magnetic order for some low temperature.

---

[2] *The result can be generalized for any value $n_i \leq 0.35$.*

On the other hand, to take into account the effect of size, we consider two lattice size 2×2×2 primitive unit cell (319 Zn atoms) and 4×4×4 supercell (1270 Zn atoms). We plot the temperature dependence of the susceptibility introduced previously and the specific heat $C_V \approx \frac{1}{T^2}\left(\langle E^2 \rangle - \langle E \rangle^2\right)$ for different cell sizes of the system $Zn_{0.7}Mn_{0.3}O$ at $n_c=0.18$.

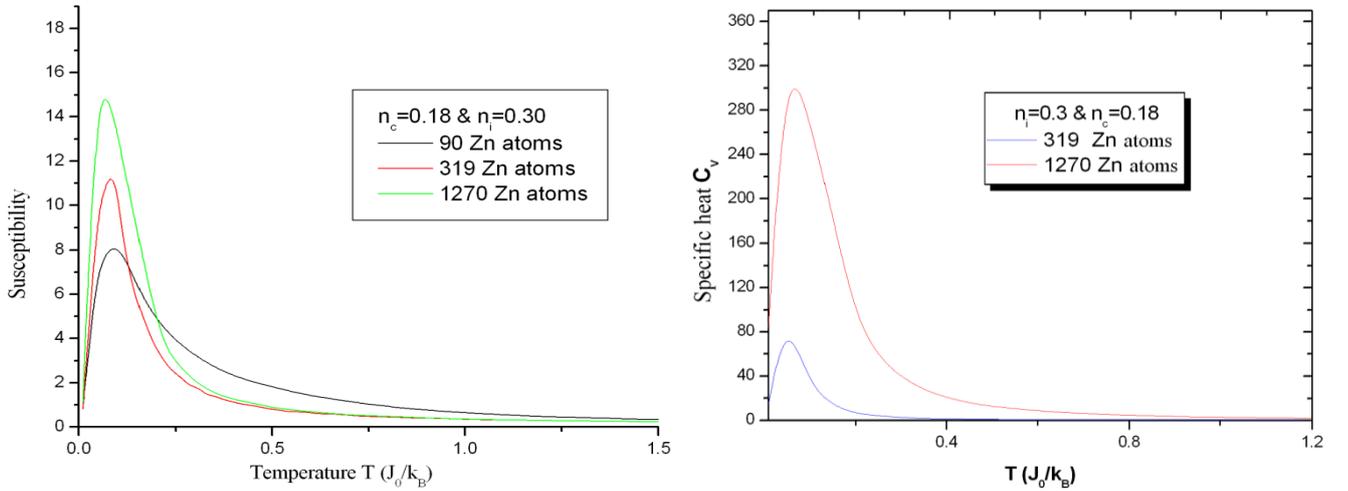

***Figure 7:*** At fixed nc=0.18 and ni=0.3 and for different cell size, **(a)** Scusceptibility versus Temperature. **(b)** Specific heat $C_v$ as a function of temperature

In figure 7a, the three curves depicting magnetic susceptibility for 1×1×1, 2×2×2 and 4×4×4 primitive peak for different values of T. In figure 7b the two curves corresponding to 2×2×2 primitive unit cell and 4×4×4 supercell peaks at specific value of temperature T and then drops to vanish. We observe that the specific heat is large for the biggest cell (1270 atoms), and becomes smaller as the size decreases.

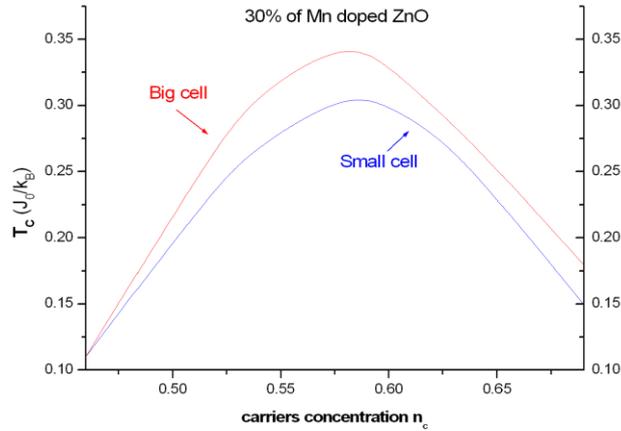

**Figure 8:** *Temperature vs holes density for 30% of Mn diluted in ZnO containing 319 Zn atoms (small cell) and 1270 Zn atoms (big cell).*

Motivated by the subsequent discussion and because of the interesting results got in the region III , we restrict on $n_c$ varying in the range from 0.42 to 0.75. As shown in figure 8, the two curves of the temperature $T_C$ increases with $n_c$ and reach their maximum at 0.58; then they sharply drop as a function of (–nc) for higher values. This maximum increases as the cell size is bigger which agrees very well with literature.

## Conclusion

In this paper, we have studied $Zn_{1-x}Mn_xO$ with concentration $n_i$ of Mn dopant ranging from 0.15 to 0.35 and for different values of the carrier concentrations $n_c$. The Monte Carlo simulations are performed to determine $T_C$ of those systems. We find that the thought controversy concerning magnetic order can be understood on the basis of investigating the effect of carriers in the system. The present study offers a specific example of how carriers can induce and control magnetic order.

So from this study we learn that $T_C$ is function of many parameters as magnetic cation concentrations $n_i$ , cell size, short or long range, holes density $n_c$. This last parameter is very crucial because as we have showed the region III -that corresponds to the case where three neighbors are FM- gives the highest value of $T_C$, however the critical temperature has lower values in the region I where the first and second near neighbors are ferromagnetic but the third one is antiferromagnetic. So this work showed us that, for the case of $Zn_{1-x}Mn_xO$, to get the FM phase with high $T_C$, the concentration of carriers should belong to the range [0.52-0.65]. And we give explicitly the regions of the values of $n_c$ that must be avoided.

## Comments and Discussions

Several models have been proposed to explain the nature of the exchange interaction in DMS; double exchange, Bound magnetic polaron and Ruderman-Kittel-Kasuya-Yosida (RKKY) interaction. The ab intio calculation of doped ZnO with defects, show that a dominant interaction is RKKY [33,36]. It belongs to one of the most important and frequently discussed couplings between the localized magnetic moments in solids. Where indirect exchange coupling is applied to localize inner d-electron of transition metal spins via conduction electrons. Such a coupling between the distant magnetic moments can also be a source of decoherence [37]. Our results here show that the transition temperature depends on the magnetic cation concentrations $n_i$, and holes density $n_c$. For low $n_c$ , the region where the first and second near neighbors are ferromagnetic and the third one is antiferromagnetic, the system exhibit low transition temperature. For intermediate value of $n_c$ concentration, where the frustration is dominant, the system is paramagnetic. While for enough high $n_c$ concentration, one obtain high transition temperature. This confirms that in order to stabilize the FM phase in $Zn_{1-x}Mn_xO$ system, it is necessary to insert carriers with concentration $n_c$ belonging to the regions I or much better to region III. So, the various experimental and theoretical investigations of the magnetic order in $Zn_{1-x}Mn_xO$ doesn't give contradictory results as now we can explain the ferromagnetism in (Zn,Mn)O systems reported in [2] or the anti-ferromagnetic or spin-glass behavior observed in [12], [17].

**Acknowledgements:** *Lalla Btissam Drissi would like to thank ICTP (Trieste) for the junior associateship scheme. L.B. Drissi would like also to thank P. Ghosh for discussions.*